\documentclass[preprint,draft,prl,10pt,twocolumn]{revtex4}

\input{tcilatex}
\usepackage{graphicx}

\begin{document}

\title{Heralded generation of multi-photon entanglement}

\author{Philip Walther$^{1,\dagger}$, Markus Aspelmeyer$^{1,2}$, Anton Zeilinger$^{1,2}$}
 \affiliation{$^1$~Institute for Experimental Physics, University of Vienna, Boltzmanngasse 5, A-1090 Vienna, Austria\\
 $^2$~Institute for Quantum Optics and Quantum Information (IQOQI), Austrian Academy of Sciences, Boltzmanngasse 3, A-1090 Vienna, Austria\\
 $^{\dagger}$~present address: Department of Physics, Harvard University, 17 Oxford Street, Cambridge, MA 02138, USA}

\begin{abstract}
We present a new scheme, based only on linear optics and standard
photon detection, that allows to generate heralded multi-photon
entangled states of arbitrary photon number from spontaneous
parametric downconversion (PDC) in the weak interaction regime.
The scheme also works in the strong interaction regime, i.e. for
the production of large photon numbers, when photon-number
resolving detectors with nearly perfect quantum efficiency are
used. In addition, the same setup can be used for quantum
metrology as multi-photon interferometer with sensitivity at the
Heisenberg-limit.
\end{abstract}

\maketitle

The controlled generation of single photons and of multi-photon
entangled states is at the heart of quantum information
processing, in particular quantum
cryptography~\cite{Bennett84,Gisin02} and scalable approaches
towards photonics-based quantum computing
schemes~\cite{Knill00,Gottesman99,Browne04}. Up to now the best
source for heralded single-photon states is spontaneous parametric
down-conversion (PDC)~\cite{Burnham70}, in which pairs of
strongly time-correlated photons are emitted into two spatial
modes and where the detection of one photon of a pair indicates
with a high probability the presence of a single-photon in the
second mode. Such conditional methods achieve preparation
efficiencies of up to 85\%~\cite{Uren_2004} with state qualities
of $g^{(2)}(0) \approx 1\times10^{-3}$, yet unchallenged by
alternative approaches based on atomic or solid states systems.
An important goal is to achieve similar heralding for entangled
multi-photon states. A problem arises because of the
probabilistic emission of PDC. Since pairs of photons are not
created 'event-ready'~\cite{Zukowski93}, \textit{all} photons
involved in a protocol need to be measured to ensure that the
wanted multi-photon state has been created. Several approaches
exist to overcome the probabilistic emission in PDC and to
prepare two-photon entangled states conditioned on the detection
of auxiliary
photons~\cite{kok_2000,Sliwa_2003,IEEE2003,Hnilo2005,Eisenberg2005}.
However, these suggested schemes have either low production
probability at high state fidelities for their generated
entangled pairs, or they rely on still unavailable detectors that
unambiguously resolve photon numbers or on nonlinear interactions
between photons, which are too weak to be exploited for
applications. We propose a new scheme to generate heralded
multi-photon states of arbitrary photon number using only linear
optics and standard (non number-resolving) photon detectors. We
suggest the heralded generation of N-particle entangled states of
the GHZ type, $|\Psi\rangle^{(GHZ)}=1/\sqrt{2}(|0\rangle^{\otimes
N}+|1\rangle^{\otimes N})$, from a 2N pair-emission from PDC. In
addition, the setup can be used to generate path-entangled
photon-number states of the NOON type,
$|\Psi\rangle^{(NOON)}=1/\sqrt(2)(|4N\rangle|0\rangle+|0\rangle|4N\rangle)$,
which are of practical relevance for quantum metrology at the
Heisenberg limit.

Our proposal is based on separating photon pairs into different
pairs of modes and utilizing two-particle-interferometry rather
than distinguishing photon numbers or employing nonlinear
beamsplitters. A crucial element is the polarizing beam splitter
(PBS) that transmits horizontally ($H$) polarized light and
reflects vertically ($V$) polarized light. This property of the
PBS allows 'targeted projections' onto either path-entangled
states~\cite{Walther04a} or polarization-entangled
states~\cite{Pan_2001} or a combination of both. Such a successful
two-photon interaction can be detected by a two-photon coincidence
measurement, where in each output mode one and only one photon is
propagating. For the case where two photons arrive from different
input modes the PBS acts as parity check~\cite{Pittman01}. If the
input photons propagate within the same mode they must have
orthogonal polarizations for being split up into the two
different output modes.

Fig. 1 gives a schematic diagram of a possible setup to generate
N-particle quantum states of the GHZ type non-destructively, i.e.
event-ready. Consider first the case for the heralded generation
of the Bell state state
$|\Phi^+\rangle=1/\sqrt{2}(|0\rangle^{\otimes
2}+|1\rangle^{\otimes 2})$ (Fig. 1a). An ultra-violet (UV) pulse
passes through two co-linearly aligned crystals for PDC,
probabilistically producing pairs of energy-degenerate
polarization-entangled photons into the spatial modes $a2$ \&
$b2$ and $a4$ \& $b4$. The UV pulse is reflected back at an
adjustable mirror, and can thus also emit entangled photon pairs
into the modes $a1$ \& $b1$ and $a3$ \& $b3$. The setup is aligned
to produce the Bell state
\begin{equation}
|\Phi^+\rangle=1/\sqrt{2}(|H\rangle_a|H\rangle_b+|V\rangle_a|V\rangle_b)
\end{equation}
for each of the pairs generated into the pairs of modes $a1$ \&
$b1$, $a2$ \& $b2$, etc. If the back-reflection mirror is kept
interferometrically stable the emission amplitudes add up
coherently and the overall Hamiltonian for creating 4 pairs via
PDC can be written as
\begin{eqnarray}
\mathcal{H}\propto [\sum_{m=2,4}(a^{\dagger}m_H b^{\dagger}m_H+
a^{\dagger}m_V b^{\dagger}m_V)+ \nonumber \\
\sum_{n=1,3}e^{i\Delta\phi}(a^{\dagger}n_H b^{\dagger}n_H+
a^{\dagger}n_V b^{\dagger}n_V)]^4 + h.c.
\end{eqnarray}
Here, $a^{\dagger}m$ and $b^{\dagger}m$ are creation operators
for the forward emitted entangled photon pairs ($a2$ \& $b2$ and
$a4$ \& $b4$), and $a^{\dagger}n$ and $b^{\dagger}n$ are the
creation operators for the backward emitted entangled
photon-pairs ($a1$ \& $b1$ and $a3$ \& $b3$). The phase of each
backward emitted photon pair is influenced by the position $x$ of
the pump mirror PM with $\Delta\phi=2\pi\frac{\Delta x}{\lambda}$
and the single-photon wavelength $\lambda$ of the downconverted
photons.

To see the working mechanism of the scheme the setup can be split
up in two parts (Fig. 1): on the left side the PBSs combine
forward- and backward emission modes from different crystals
($a2$ \& $a3$ and $a4$ \& $a1$) and on the right side forward-
and backward emission modes from the same crystals are
superimposed ($b1$ \& $b2$ and $b3$ \& $b4$). Assuming that the
probability for the higher order emission of 5 and more photon
pairs is negligible (this is typically the case in the weak
interaction regime of PDC; see below), only three different
emission processes can lead to the propagation of 8 photons in
the 8 output modes behind the beamsplitters: either one entangled
pair is emitted into each of the modes $a1$ \& $b1$, $a2$ \& $b2$,
$a3$ \& $b3$, $a4$ \& $b4$ or a double-pair emission occurs into
the backward modes both $a1$ \& $b1$ and $a3$ \& $b3$ or a
double-pair emission occurs into the forward modes both $a2$ \&
$b2$ and $a4$ \& $b4$. Because of the specific choice of modes to
overlap at the PBSs all other possible emissions fail to provide
sufficiently many photons in all output modes (For example, a
three-pair emission into modes $a2$ \& $b2$ together with a
single-pair emission into modes $a4$ \& $b4$ would fail to
provide a coincident detection event in the output modes $a1'$ \&
$a4'$). The left side acts as a parity check on the photon
polarization in emission modes $a2$ \& $a3$ and $a4$ \& $a1$.
Because of the polarization-entanglement of the initially emitted
pairs the overall state is projected into
\begin{eqnarray}
&\hspace{-2mm}|V\rangle_{a1'}|H\rangle_{a2'}|V\rangle_{a3'}|H\rangle_{a4'}|V\rangle_{b1'}|H\rangle_{b2'}|V\rangle_{b3'}|H\rangle_{b4'}
+  \nonumber \\
e^{i8\Delta\phi}&\hspace{-2mm}|H\rangle_{a1'}\hspace{-1mm}|V\rangle_{a2'}|H\rangle_{a3'}|V\rangle_{a4'}|H\rangle_{b1'}|V\rangle_{b2'}|H\rangle_{b3'}|V\rangle_{b4'}+
\nonumber \\
e^{i4\Delta\phi}&\hspace{-2mm}(|H\rangle_{a1'}|H\rangle_{a2'}|H\rangle_{a3'}|H\rangle_{a4'}|H\rangle_{b1'}|H\rangle_{b2'}|H\rangle_{b3'}|H\rangle_{b4'}+\nonumber \\
&\hspace{-2mm}|V\rangle_{a1'}|V\rangle_{a2'}|V\rangle_{a3'}|V\rangle_{a4'}|V\rangle_{b1'}|V\rangle_{b2'}|V\rangle_{b3'}|V\rangle_{b4'}+
\nonumber\\
&\hspace{-2mm}|H\rangle_{a1'}|V\rangle_{a2'}|V\rangle_{a3'}|H\rangle_{a4'}|HV\rangle_{b1'}|0\rangle_{b2'}|0\rangle_{b3'}|HV\rangle_{b4'}+
\nonumber\\
&\hspace{-2mm}|V\rangle_{a1'}|H\rangle_{a2'}|H\rangle_{a3'}|V\rangle_{a4'}|0\rangle_{b1'}|VH\rangle_{b2'}|VH\rangle_{b3'}|0\rangle_{b4'}).
\label{after_parity}
\end{eqnarray}

The first term stems from the two double-pair emissions into the
forward modes $a2$ \& $b2$ and $a4$ \& $b4$ and hence carries a
phase of $e^{i8\Delta\phi}$ relative to the second term that is
due to the two double-pair emissions into the backward modes $a1$
\& $b1$ and $a3$ \& $b3$. The third term, with a relative phase of
$e^{i4\Delta\phi}$, originates from the single-pair emission into
all mode-pairs. Note that it has two contributions in which no
photons are emitted into modes $b1'$ or $b3'$. The parity check
leaves therefore only those contributions in which a detection
event in $b1'$ and $b3'$ indicates \textit{with certainty} that a
photon each propagates in modes $b2'$ and $b4'$. In this way, the
emission of a two-photon state in modes $b2'$ and $b4'$ is
heralded by joint detection events in modes
$a1',a2',a3',a4',b1'$and $b3'$. By projecting each of these 6
modes into a specific linearly polarized basis, e.g.
$|\pm\rangle=\frac{1}{\sqrt{2}}(|H\rangle \pm|V\rangle)$, one
obtains the maximally entangled Bell state
$|\Phi^+\rangle=\frac{1}{\sqrt{2}}(|H\rangle_{b2'}|H\rangle_{b4'}+|V\rangle_{b2'}|V\rangle_{b4'})$,
as can be seen by projecting the state of Eq.~(\ref{after_parity})
onto $|+\rangle_{a1'} |+\rangle_{a2'} |+\rangle_{a3'}
|+\rangle_{a4'} |+\rangle_{b1'} |+\rangle_{b3'}$. This entangled
state, $|\Phi^+\rangle$, can be easily converted into any other
Bell state by using local operations at one of the qubits. Recent
6-photon experiments were already able to detect 6-fold
coincidences at a rate of 40 per minute and with 2-photon
visibilities beyond 90\%~\cite{Lu2006}, which corresponds
respectively to the expected generation rate and fidelity of the
heralded pair if the same parameters were used. It is also
important to note that our scheme does not need photon-number
resolution in the detection process, since the scheme inherently
suppresses situations in which more than one photon is emitted
into each of the 6 detection modes. Furthermore, the fidelity of
the created state is independent of the detection efficiency,
which is an advantage over other proposals where this is not the
case~\cite{Uren_2004}. In our case, the fidelity will depend on
the interaction parameter. Increasing the interaction will
enhance the effect of stimulated emission and lead to spurious
detection events that will eventually diminish the fidelity. We
discuss a workaround scheme at the end of the paper. For the
present discussion we focus on the situation, in which
higher-order pair emissions are essentially negligible.

This scheme can easily be generalized to the heralded generation
of higher photon-number entangled states. By adding additional
crystals (and hence more pairs of emission modes) and by keeping
the general structure (of parity-checks on the left side and
projection on the right side) the above arguments hold as well.
For example, one additional crystal extends the scheme to 6 pairs
of emission modes. Again, the only relevant emissions are those
three where (1) one pair each is emitted into the pairs of modes
$a1\&b1$, $a2\&b2$, $a3\&b3$, $a4\&b4$, $a5\&b5$, $a6\&b6$, or
(2) a double pair each is emitted into the forward pairs of modes
$a2\&b2$, $a4\&b4$, $a6\&b6$ or (3) a double pair each is emitted
into the backward pairs of modes $a1\&b1$, $a3\&b3$, $a5\&b5$. The
left-side parity check leaves the state
\begin{eqnarray}
&|V\rangle_{a1'}|H\rangle_{a2'}|V\rangle_{a3'}|H\rangle_{a4'}|V\rangle_{a5'}|H\rangle_{a6'}\nonumber\\
&|V\rangle_{b1'}|H\rangle_{b2'}|V\rangle_{b3'}|H\rangle_{b4'}|V\rangle_{b5'}|H\rangle_{b6'}
&+ \nonumber \\
e^{i12\Delta\phi}&|H\rangle_{a1'}|V\rangle_{a2'}|H\rangle_{a3'}|V\rangle_{a4'}|H\rangle_{a5'}|V\rangle_{a6'}\nonumber\\
&|H\rangle_{b1'}|V\rangle_{b2'}|H\rangle_{b3'}|V\rangle_{b4'}|H\rangle_{b5'}|V\rangle_{b6'}&+
\nonumber \\
e^{i6\Delta\phi}(&|H\rangle_{a1'}|H\rangle_{a2'}|H\rangle_{a3'}|H\rangle_{a4'}|H\rangle_{a5'}|H\rangle_{a6'}\nonumber\\
&|H\rangle_{b1'}|H\rangle_{b2'}|H\rangle_{b3'}|H\rangle_{b4'}|H\rangle_{b5'}|H\rangle_{b6'}&+
\nonumber \\
&|V\rangle_{a1'}|V\rangle_{a2'}|V\rangle_{a3'}|V\rangle_{a4'}|V\rangle_{a5'}|V\rangle_{a6'}\nonumber\\
&|V\rangle_{b1'}|V\rangle_{b2'}|V\rangle_{b3'}|V\rangle_{b4'}|V\rangle_{b5'}|V\rangle_{b6'}).
\label{after_parity_GHZ}
\end{eqnarray}

For reasons of clarity we have omitted those terms in which no
photons are emitted into the modes $b'$. Conditioning the right
hand side on a 3-fold detection event in modes $b1', b3', b5'$
results then with certainty in a 3-photon state in the undetected
emission modes $b2', b4', b6'$. A projection of each of the nine
detection modes $a1', a2', a3', a4', a5', a6', b1', b3', b5'$
into the linear polarization basis $|\pm\rangle$ finally
generates a freely propagating 3-particle GHZ-state. For example,
a projection onto $|+\rangle_{a1'} |+\rangle_{a2'} |+\rangle_{a3'}
|+\rangle_{a4'} |+\rangle_{a5'} |+\rangle_{a6'} |+\rangle_{b1'}
|+\rangle_{b3'} |+\rangle_{b5'}$ leads to the heralded generation
of the maximally entangled state
$|GHZ\rangle^{(3)}=\frac{1}{\sqrt{2}}(|H\rangle |H\rangle |H\rangle + |V\rangle |V\rangle |V\rangle)$.\\
It is straightforward to see that this technique, applied to 2n
emission modes (n: number of crystals), is capable of heralding
n-photon GHZ-states of the type
$|GHZ\rangle^{(n)}=\frac{1}{\sqrt{2}}(|H\rangle^{\otimes
n}+|V\rangle^{\otimes n})$ (Fig. 1c). The 2n-photon parity check
on the left side together with the n-fold detection in the odd
modes after the PBSs on the right side conditions the emission of
a n-photon state into the n remaining even modes on the left
side. If the 3n-photon coincidence detection is performed in a
specific linear polarization basis $|\pm\rangle$ the heralded
state is the wanted quantum state of the GHZ-type.

We would like to note an additional feature of the presented
scheme. If one decides to measure all 4n output ports via a
4n-fold coincidence detection, the setup represents an
interferometer with a sub shot-noise phase sensitivity. Such
performance is known in the context of quantum metrology and can
occur when the state propagating within the interferometer is
highly non-classical. For example, path-entangled photon-number
states, so-called NOON-states~\cite{Kok01}, are known to achieve
Heisenberg-limited interferometric sensitivity beyond the
classical shot-noise limit. In a regular Mach-Zehnder
interferometer, such states are of the form
$|\Psi\rangle^{(NOON)}=1/\sqrt(2)(|N\rangle_A|0\rangle_B+|0\rangle_A|N\rangle_B)$,
where $A$ and $B$ are the two spatial modes of propagation within
the interferometer. Here, we generate NOON-states that consist of
a nonlocal 4n-photon state~\cite{Walther04a} that is
path-entangled between the n pairs of modes for forward emission,
$a2\&b2$, $a4\&b4$, ... and the n pairs of modes for backward
emission, $a1\&b1$, $a3\&b3$, ..., i.e.
$|\Psi\rangle=1/\sqrt(2)(|4n\rangle_{a1\&b1,a3\&b3,...}
|0\rangle_{a2\&b2,a4\&b4,...} + |0\rangle_{a1\&b1,a3\&b3,...}
|4n\rangle_{a2\&b2,a4\&b4,...}$. Specifically, if we project all
4n detection modes onto linear polarizations $|\pm\rangle$ with an
odd number of $|-\rangle$ projections, e.g. $|-\rangle_{a1'}
|+\rangle_{b1'} |+\rangle_{a2'}
|+\rangle_{b2'}...|+\rangle_{a2n'} |+\rangle_{b2n'}$, then the
4n-fold coincidence detection probability varies with
$P_{a1',b1',...,an',bn'}\propto1-cos{4n\Delta\phi}$ and the
phase-noise sensitivity $<\Delta\phi>$ scales with $\frac{1}{4n}$
as opposed to the expected $\frac{1}{\sqrt{4n}}$ (shot-noise)
performance of a comparable "classical" interferometer (in
particular, a single-photon interferometer with 2n independent
averaging runs or an interferometer that is driven with a
coherent state with $\alpha\approx2n$). Interferometers based on
NOON-states have recently been demonstrated for up to 4
photons~\cite{Rarity90,Walther04a,Mitchell04} and up to 6
ions~\cite{Sackett00,Leibfried04,Leibfried_2005}.

As an example, consider the 8-photon interferometer described in
Figure 2a, where n=2. Projecting each of the eight output modes
onto a linear basis, $|\pm\rangle$, specifically
\begin{eqnarray}
|-\rangle_{a1'}|+\rangle_{a2'}|+\rangle_{a3'}|+\rangle_{a4'}|+\rangle_{b1'}|+\rangle_{b2'}|+\rangle_{b3'}|+\rangle_{b4'},\nonumber
\end{eqnarray}
effectively eliminates the contributions of the emission process
into each pairs of modes (3rd term in Eq.~\ref{after_parity}) and
results in the maximally path entangled state of the form
\begin{eqnarray}
|\Psi\rangle=1/\sqrt{2}(|8\rangle_{a2'b2'a4'b4'}|0\rangle_{a1'b1'a3'b3'}+
\nonumber \\
e^{i8\Delta\phi}|0\rangle_{a2'b2'a4'b4'}|8\rangle_{a1'b1'a3'b3'}).\nonumber
\end{eqnarray}

These remaining contributions correspond to a $NOON$-state, where
either 8 photons are emitted via four-photon emission from each
crystal into the pair of modes $a2$ \& $b2$ and $a4$ \& $b4$ or
by four-photon emissions into the backward pair of modes $a1$ \&
$b1$ and $a3$ \& $b3$. Thus the eight-photon detection
probability in the spatially separated output modes oscillates
like
$P_{a1',a2',a3',a4',b1',b2',b3',b4'}\propto1-cos(8\Delta\phi)$
when moving the delay mirror for the UV pump beam. Finally, the
generalization to an arbitrary number of 4n photons is equivalent
to the heralding scheme above and simply consists of adding
crystals and keeping the overall structure of the interferometric
scheme (Fig. 2b).

Let us finally comment on the potentials and limitations of the
scheme. Firstly, the setup is essentially a multi-photon
interferometer in which the phase needs to be controlled with
high accuracy. This poses an increasing challenge with increasing
number of involved photons, since the phase-sensitivity of the
state contributions scale linearly with photon number (see Eq.
(\ref{after_parity}) and (\ref{after_parity_GHZ})). However,
locking techniques can achieve stable interferometer controls to
better than $10^{-15}m$~\cite{attometer}, which would allow photon
numbers on the order of $10^8$ or larger; hence phase stability
does not pose a limitation. Secondly, the scheme is based on using
stimulated PDC emission in the weak interaction regime where the
probability to generate an additional pair is negligible. In our
case, if (2n+1) pairs would be emitted instead of the wanted 2n,
spurious coincidence events might occur without generating the
heralded state. The probability to create an additional pair is
on the order of the interaction parameter\cite{Kok_2000a}, which
is typically between 0.01 and 0.1~\cite{Kaltenbaek05}. However,
although increasing the interaction leads to larger
photon-numbers it also increases the effect of stimulated
emission~\cite{Linares_2001,Eisenberg_2004}. This eventually
results in the more efficient production of states with higher
pair numbers and hence decreases the performance of our scheme if
non photon-number resolving detectors are used. The use of
photon-number resolving detectors with almost perfect quantum
efficiency would overcome this deficiency and the scheme would
still work.

In conclusion, we have presented a scalable scheme to herald the
generation of multi-photon entangled states of, in principle,
arbitrary size. The setup can also be viewed (and used) as
Heisenberg-limited interferometer. For PDC in the weak
interaction regime standard photon detection without number
resolution is sufficient to generate high-quality GHZ-type
quantum states in an event-ready manner. For the strong
interaction regime, which is necessary to achieve large photon
numbers~\cite{Eisenberg_2004}, the use of photon-number resolving
detectors allows to maintain the proposed performance.

\begin{acknowledgments}
The authors thank Kevin Resch for helpful discussions. This work
was supported by the Austrian Science Foundation (FWF) under
project SFB15,the European Commission, contract numbers
IST-2001-38864 (RAMBOQ) and IST-015848 (QAP), the DTO-funded U.S.
Army Research Office Contract No. W911NF-05-0397 and by the
Alexander von Humboldt-Foundation.
\end{acknowledgments}


\noindent {\bf Fig. 1.} Schematic setup for the heralded
preparation of multi-photon maximally entangled states. {\bf a}
Conditional generation of two-photon Bell-states. {\bf b}
Conditional generation of three-photon GHZ-states. {\bf c}
Conditional generation of multi-photon states of the GHZ-type.

\noindent {\bf Fig. 2.} Multi-photon interferometer for quantum
metrology. {\bf a} 8-photon path-entangled NOON-states can be
generated by performing a polarization measurement in all of the
output modes in the linear polarization basis $|\pm\rangle$ where
the number of $|-\rangle$ projections is odd. This results in pure
8-photon interference due to projection onto the (non-local)
path-entangled 8-photon state
$|\Psi\rangle=1/\sqrt{2}(|8\rangle_{a2b2a4b4}
|0\rangle_{a1b1a3b3} + |0\rangle_{a2b2a4b4} |8\rangle_{a1b1a3b3})$
. {\bf b} Generalized scheme for 4n-photon NOON states.

%
%
%
%


\begin{thebibliography}{23}
\expandafter\ifx\csname
natexlab\endcsname\relax\def\natexlab#1{#1}\fi
\expandafter\ifx\csname bibnamefont\endcsname\relax
  \def\bibnamefont#1{#1}\fi
\expandafter\ifx\csname bibfnamefont\endcsname\relax
  \def\bibfnamefont#1{#1}\fi
\expandafter\ifx\csname citenamefont\endcsname\relax
  \def\citenamefont#1{#1}\fi
\expandafter\ifx\csname url\endcsname\relax
  \def\url#1{\texttt{#1}}\fi
\expandafter\ifx\csname
urlprefix\endcsname\relax\def\urlprefix{URL }\fi
\providecommand{\bibinfo}[2]{#2}
\providecommand{\eprint}[2][]{\url{#2}}

\bibitem[{\citenamefont{Bennett and Brassard}(1984)}]{Bennett84}
\bibinfo{author}{\bibfnamefont{C.~H.} \bibnamefont{Bennett}} \bibnamefont{and}
  \bibinfo{author}{\bibfnamefont{G.}~\bibnamefont{Brassard}},
  \bibinfo{journal}{Proceedings of IEEE International Conference on Computers,
  Systems and Signa Processing, Bangalore, India} pp. \bibinfo{pages}{175--179}
  (\bibinfo{year}{1984}).

\bibitem[{\citenamefont{Gisin et~al.}(2002)\citenamefont{Gisin, Ribordy,
  Tittel, and Zbinden}}]{Gisin02}
\bibinfo{author}{\bibfnamefont{N.}~\bibnamefont{Gisin}},
  \bibinfo{author}{\bibfnamefont{G.}~\bibnamefont{Ribordy}},
  \bibinfo{author}{\bibfnamefont{W.}~\bibnamefont{Tittel}}, \bibnamefont{and}
  \bibinfo{author}{\bibfnamefont{H.}~\bibnamefont{Zbinden}},
  \bibinfo{journal}{Rev.\ Mod.\ Phys.} \textbf{\bibinfo{volume}{74}},
  \bibinfo{pages}{145} (\bibinfo{year}{2002}).

\bibitem[{\citenamefont{Knill et~al.}(2000)\citenamefont{Knill, Laflamme, and
  Milburn}}]{Knill00}
\bibinfo{author}{\bibfnamefont{E.}~\bibnamefont{Knill}},
  \bibinfo{author}{\bibfnamefont{R.}~\bibnamefont{Laflamme}}, \bibnamefont{and}
  \bibinfo{author}{\bibfnamefont{G.}~\bibnamefont{Milburn}},
  \bibinfo{journal}{Nature} \textbf{\bibinfo{volume}{409}}, \bibinfo{pages}{46}
  (\bibinfo{year}{2000}).

\bibitem[{\citenamefont{Gottesman and Chuang}(1999)}]{Gottesman99}
\bibinfo{author}{\bibfnamefont{D.}~\bibnamefont{Gottesman}} \bibnamefont{and}
  \bibinfo{author}{\bibfnamefont{I.~L.} \bibnamefont{Chuang}},
  \bibinfo{journal}{Nature} \textbf{\bibinfo{volume}{402}},
  \bibinfo{pages}{390} (\bibinfo{year}{1999}).

\bibitem[{\citenamefont{Browne and Rudolph}(2005)}]{Browne04}
\bibinfo{author}{\bibfnamefont{D.~E.} \bibnamefont{Browne}} \bibnamefont{and}
  \bibinfo{author}{\bibfnamefont{T.}~\bibnamefont{Rudolph}},
  \bibinfo{journal}{Phys.\ Rev.\ Lett.} \textbf{\bibinfo{volume}{95}},
  \bibinfo{pages}{010501} (\bibinfo{year}{2005}).

\bibitem[{\citenamefont{Burnham and Weinberg}(1970)}]{Burnham70}
\bibinfo{author}{\bibfnamefont{D.~C.} \bibnamefont{Burnham}} \bibnamefont{and}
  \bibinfo{author}{\bibfnamefont{D.~L.} \bibnamefont{Weinberg}},
  \bibinfo{journal}{Phys.\ Rev.\ Lett.} \textbf{\bibinfo{volume}{25}},
  \bibinfo{pages}{84} (\bibinfo{year}{1970}).

\bibitem[{\citenamefont{Lu et~al.}(2006)\citenamefont{Lu, Zhou, G\"uhne, Gao, Zhang,
Yuan, Goebel, Yang, and Pan}}]{Lu2006}
\bibinfo{author}{\bibfnamefont{C.-Y.} \bibnamefont{Lu}},
  \bibinfo{author}{\bibfnamefont{X.-Q.}~\bibnamefont{Zhou}},
  \bibinfo{author}{\bibfnamefont{O.}~\bibnamefont{G\"uhne}},
  \bibinfo{author}{\bibfnamefont{W.-B.}~\bibnamefont{Gao}},
  \bibinfo{author}{\bibfnamefont{J.}~\bibnamefont{Zhang}},
  \bibinfo{author}{\bibfnamefont{Z.-S.}~\bibnamefont{Yuan}},
  \bibinfo{author}{\bibfnamefont{A.}~\bibnamefont{Goebel}},
  \bibinfo{author}{\bibfnamefont{T.}~\bibnamefont{Yang}},
  \bibinfo{author}{\bibfnamefont{J.-W.}~\bibnamefont{Pan}},
  \bibinfo{journal}{quant-ph/0610145}
  (\bibinfo{year}{2006}).


\bibitem[{\citenamefont{U'Ren et~al.}(2004)\citenamefont{U'Ren, Silberhorn,
  Banaszek, and A.Walmsley}}]{Uren_2004}
\bibinfo{author}{\bibfnamefont{A.~B.} \bibnamefont{URen}},
  \bibinfo{author}{\bibfnamefont{C.}~\bibnamefont{Silberhorn}},
  \bibinfo{author}{\bibfnamefont{K.}~\bibnamefont{Banaszek}}, \bibnamefont{and}
  \bibinfo{author}{\bibfnamefont{I.}~\bibnamefont{A.Walmsley}},
  \bibinfo{journal}{Phys.\ Rev.\ Lett.}
  \textbf{\bibinfo{volume}{93}},  \bibinfo{pages}{093601}
  (\bibinfo{year}{2004}).

\bibitem[{\citenamefont{Zukowski et~al.}(1993)\citenamefont{Zukowski,
  Zeilinger, Horne, and Ekert}}]{Zukowski93}
\bibinfo{author}{\bibfnamefont{M.}~\bibnamefont{Zukowski}},
  \bibinfo{author}{\bibfnamefont{A.}~\bibnamefont{Zeilinger}},
  \bibinfo{author}{\bibfnamefont{M.~A.} \bibnamefont{Horne}}, \bibnamefont{and}
  \bibinfo{author}{\bibfnamefont{A.~K.} \bibnamefont{Ekert}},
  \bibinfo{journal}{Phys.\ Rev.\ Lett} \textbf{\bibinfo{volume}{71}},
  \bibinfo{pages}{4287} (\bibinfo{year}{1993}).

\bibitem[{\citenamefont{Kok}(2000)}]{kok_2000}
\bibinfo{author}{\bibfnamefont{P.}~\bibnamefont{Kok}}, Ph.D. thesis,
  \bibinfo{school}{University of Wales, Bangor} (\bibinfo{year}{2000}).

\bibitem[{\citenamefont{Sliwa and Banaszek}(2003)}]{Sliwa_2003}
\bibinfo{author}{\bibfnamefont{C.}~\bibnamefont{Sliwa}} \bibnamefont{and}
  \bibinfo{author}{\bibfnamefont{K.}~\bibnamefont{Banaszek}},
  \bibinfo{journal}{Phys.\ Rev.\ A} \textbf{\bibinfo{volume}{67}},
  \bibinfo{pages}{030101(R)} (\bibinfo{year}{2003}).

\bibitem{IEEE2003} T.B. Pittman et al., \emph{IEEE J. of Selected Topics in Quant. Elect.}
\textbf{9}, 1478 (2003).

\bibitem{Hnilo2005} A.A. Hnilo, \emph{Phys. Rev. A} \textbf{71}, 033820 (2005).

\bibitem{Eisenberg2005} H.S. Eisenberg, J.F. Hodelin, G. Khoury, and D. Bouwmeester,
\emph{Phys. Rev. Lett.} \textbf{94}, 090502(2005).

\bibitem[{\citenamefont{Walther et~al.}(2004)\citenamefont{Walther, Pan,
  Aspelmeyer, Ursin, Gasparon, and Zeilinger}}]{Walther04a}
\bibinfo{author}{\bibfnamefont{P.}~\bibnamefont{Walther}},
  \bibinfo{author}{\bibfnamefont{J.-W.} \bibnamefont{Pan}},
  \bibinfo{author}{\bibfnamefont{M.}~\bibnamefont{Aspelmeyer}},
  \bibinfo{author}{\bibfnamefont{R.}~\bibnamefont{Ursin}},
  \bibinfo{author}{\bibfnamefont{S.}~\bibnamefont{Gasparon}}, \bibnamefont{and}
  \bibinfo{author}{\bibfnamefont{A.}~\bibnamefont{Zeilinger}},
  \bibinfo{journal}{Nature} \textbf{\bibinfo{volume}{429}},
  \bibinfo{pages}{158} (\bibinfo{year}{2004}).

\bibitem[{\citenamefont{Pan et~al.}(2001)\citenamefont{Pan, Daniell, Gasparoni,
  Weihs, and Zeilinger}}]{Pan_2001}
\bibinfo{author}{\bibfnamefont{J.-W.} \bibnamefont{Pan}},
  \bibinfo{author}{\bibfnamefont{M.}~\bibnamefont{Daniell}},
  \bibinfo{author}{\bibfnamefont{S.}~\bibnamefont{Gasparoni}},
  \bibinfo{author}{\bibfnamefont{G.}~\bibnamefont{Weihs}}, \bibnamefont{and}
  \bibinfo{author}{\bibfnamefont{A.}~\bibnamefont{Zeilinger}},
  \bibinfo{journal}{Phys.\ Rev.\ Lett.} \textbf{\bibinfo{volume}{86}},
  \bibinfo{pages}{4435} (\bibinfo{year}{2001}).

\bibitem[{\citenamefont{Pittman et~al.}(2001)\citenamefont{Pittman, Jacobs, and
  Franson}}]{Pittman01}
\bibinfo{author}{\bibfnamefont{T.~B.}~\bibnamefont{Pittman}},
  \bibinfo{author}{\bibfnamefont{B.~C.}~\bibnamefont{Jacobs}}, \bibnamefont{and}
  \bibinfo{author}{\bibfnamefont{J.~D.}~\bibnamefont{Franson}},
  \bibinfo{journal}{Phys.\ Rev.\ A} \textbf{\bibinfo{volume}{64}},
  \bibinfo{pages}{062311} (\bibinfo{year}{2001}).

\bibitem[{\citenamefont{Kok}(2001)}]{Kok01}
\bibinfo{author}{\bibfnamefont{P.}~\bibnamefont{Kok}}, \bibinfo{journal}{Phys.\
  Rev.\ A} \textbf{\bibinfo{volume}{63}}, \bibinfo{pages}{063407}
  (\bibinfo{year}{2001}).

\bibitem[{\citenamefont{Rarity et~al.}(1990)\citenamefont{Rarity, Tapster,
  Jakeman, Larchuk, Campos, Teich, and Saleh}}]{Rarity90}
\bibinfo{author}{\bibfnamefont{J.~G.} \bibnamefont{Rarity}},
  \bibinfo{author}{\bibfnamefont{P.~R.} \bibnamefont{Tapster}},
  \bibinfo{author}{\bibfnamefont{E.}~\bibnamefont{Jakeman}},
  \bibinfo{author}{\bibfnamefont{T.}~\bibnamefont{Larchuk}},
  \bibinfo{author}{\bibfnamefont{R.~A.} \bibnamefont{Campos}},
  \bibinfo{author}{\bibfnamefont{M.~C.} \bibnamefont{Teich}}, \bibnamefont{and}
  \bibinfo{author}{\bibfnamefont{B.~E.~A.} \bibnamefont{Saleh}},
  \bibinfo{journal}{Phys.\ Rev.\ Lett.} \textbf{\bibinfo{volume}{65}},
  \bibinfo{pages}{1348} (\bibinfo{year}{1990}).

\bibitem[{\citenamefont{Mitchell et~al.}(2004)\citenamefont{Mitchell, Lundeen,
  and Steinberg}}]{Mitchell04}
\bibinfo{author}{\bibfnamefont{M.~W.} \bibnamefont{Mitchell}},
  \bibinfo{author}{\bibfnamefont{J.~S.} \bibnamefont{Lundeen}},
  \bibnamefont{and} \bibinfo{author}{\bibfnamefont{A.~M.}
  \bibnamefont{Steinberg}}, \bibinfo{journal}{Nature}
  \textbf{\bibinfo{volume}{429}}, \bibinfo{pages}{161} (\bibinfo{year}{2004}).

\bibitem[{\citenamefont{Sackett et~al.}(2000)\citenamefont{Sackett, Kielpinski,
  King, Langer, Meye, Myatt, Rowe, Turchette, Itano, Wineland
  et~al.}}]{Sackett00}
\bibinfo{author}{\bibfnamefont{C.~A.} \bibnamefont{Sackett}},
  \bibinfo{author}{\bibfnamefont{D.}~\bibnamefont{Kielpinski}},
  \bibinfo{author}{\bibfnamefont{B.~E.} \bibnamefont{King}},
  \bibinfo{author}{\bibfnamefont{C.}~\bibnamefont{Langer}},
  \bibinfo{author}{\bibfnamefont{V.}~\bibnamefont{Meye}},
  \bibinfo{author}{\bibfnamefont{C.~J.} \bibnamefont{Myatt}},
  \bibinfo{author}{\bibfnamefont{M.}~\bibnamefont{Rowe}},
  \bibinfo{author}{\bibfnamefont{Q.~A.} \bibnamefont{Turchette}},
  \bibinfo{author}{\bibfnamefont{W.~M.} \bibnamefont{Itano}},
  \bibinfo{author}{\bibfnamefont{D.~J.} \bibnamefont{Wineland}},
  \bibnamefont{et~al.}, \bibinfo{journal}{Nature}
  \textbf{\bibinfo{volume}{404}}, \bibinfo{pages}{256} (\bibinfo{year}{2000}).

\bibitem[{\citenamefont{Leibfried et~al.}(2004)\citenamefont{Leibfried,
  Barrett, Schaetz, Britton, Chiaverin, Itano, Jost, Langer, and
  Wineland}}]{Leibfried04}
\bibinfo{author}{\bibfnamefont{D.}~\bibnamefont{Leibfried}},
  \bibinfo{author}{\bibfnamefont{M.~D.} \bibnamefont{Barrett}},
  \bibinfo{author}{\bibfnamefont{T.}~\bibnamefont{Schaetz}},
  \bibinfo{author}{\bibfnamefont{J.}~\bibnamefont{Britton}},
  \bibinfo{author}{\bibfnamefont{J.}~\bibnamefont{Chiaverin}},
  \bibinfo{author}{\bibfnamefont{W.~M.} \bibnamefont{Itano}},
  \bibinfo{author}{\bibfnamefont{J.~D.} \bibnamefont{Jost}},
  \bibinfo{author}{\bibfnamefont{C.}~\bibnamefont{Langer}}, \bibnamefont{and}
  \bibinfo{author}{\bibfnamefont{D.~J.} \bibnamefont{Wineland}},
  \bibinfo{journal}{Science} \textbf{\bibinfo{volume}{304}},
  \bibinfo{pages}{1476} (\bibinfo{year}{2004}).

\bibitem[{\citenamefont{Leibfried et~al.}(2005)\citenamefont{Leibfried, Knill,
  Seidelin, Britton, Blakestad, Chiaverini, Hume, Itano, Jost, Langer
  et~al.}}]{Leibfried_2005}
\bibinfo{author}{\bibfnamefont{D.}~\bibnamefont{Leibfried}},
  \bibinfo{author}{\bibfnamefont{E.}~\bibnamefont{Knill}},
  \bibinfo{author}{\bibfnamefont{S.}~\bibnamefont{Seidelin}},
  \bibinfo{author}{\bibfnamefont{J.}~\bibnamefont{Britton}},
  \bibinfo{author}{\bibfnamefont{R.~B.} \bibnamefont{Blakestad}},
  \bibinfo{author}{\bibfnamefont{J.}~\bibnamefont{Chiaverini}},
  \bibinfo{author}{\bibfnamefont{D.~B.} \bibnamefont{Hume}},
  \bibinfo{author}{\bibfnamefont{W.~M.} \bibnamefont{Itano}},
  \bibinfo{author}{\bibfnamefont{J.~D.} \bibnamefont{Jost}},
  \bibinfo{author}{\bibfnamefont{C.}~\bibnamefont{Langer}},
  \bibnamefont{et~al.}, \bibinfo{journal}{Nature}
  \textbf{\bibinfo{volume}{438}}, \bibinfo{pages}{639} (\bibinfo{year}{2005}).


\bibitem[{\citenamefont{Briant et~al.}(2003)\citenamefont{Briant, Cohadon, Heidmann
and Pinard}}]{attometer}
\bibinfo{author}{\bibfnamefont{T.}~\bibnamefont{Briant}},
\bibinfo{author}{\bibfnamefont{P.-F.} \bibnamefont{Cohadon}},
\bibinfo{author}{\bibfnamefont{A.} \bibnamefont{Heidmann}},
\bibinfo{author}{\bibfnamefont{M.} \bibnamefont{Pinard}},
  \bibinfo{journal}{Proc. EQEC~03},
  \bibinfo{pages}{327} (\bibinfo{year}{2003}).

\bibitem[{\citenamefont{Kok and Braunstein}(2000)}]{Kok_2000a}
\bibinfo{author}{\bibfnamefont{P.}~\bibnamefont{Kok}} \bibnamefont{and}
  \bibinfo{author}{\bibfnamefont{S.~L.} \bibnamefont{Braunstein}},
  \bibinfo{journal}{Phys.\ Rev.\ A} \textbf{\bibinfo{volume}{61}},
  \bibinfo{pages}{042304} (\bibinfo{year}{2000}).

\bibitem[{\citenamefont{Kaltenbaek et~al.}(2005)\citenamefont{Kaltenbaek,
  Blauensteiner, Zukowski, Aspelmeyer, and Zeilinger}}]{Kaltenbaek05}
\bibinfo{author}{\bibfnamefont{R.}~\bibnamefont{Kaltenbaek}},
  \bibinfo{author}{\bibfnamefont{B.}~\bibnamefont{Blauensteiner}},
  \bibinfo{author}{\bibfnamefont{M.}~\bibnamefont{Zukowski}},
  \bibinfo{author}{\bibfnamefont{M.}~\bibnamefont{Aspelmeyer}},
  \bibnamefont{and}
  \bibinfo{author}{\bibfnamefont{A.}~\bibnamefont{Zeilinger}},
  \bibinfo{journal}{to be published}  (\bibinfo{year}{2005}).

\bibitem[{\citenamefont{Lamas-Linares et~al.}(2001)\citenamefont{Lamas-Linares,
  Howell, and Bouwmeester}}]{Linares_2001}
\bibinfo{author}{\bibfnamefont{A.}~\bibnamefont{Lamas-Linares}},
  \bibinfo{author}{\bibfnamefont{J.~C.} \bibnamefont{Howell}},
  \bibnamefont{and}
  \bibinfo{author}{\bibfnamefont{D.}~\bibnamefont{Bouwmeester}},
  \bibinfo{journal}{Nature} \textbf{\bibinfo{volume}{412}},
  \bibinfo{pages}{887} (\bibinfo{year}{2001}).

\bibitem[{\citenamefont{Eisenberg et~al.}(2004)\citenamefont{Eisenberg, Khoury,
  Durkin, Simon, and Bouwmeester}}]{Eisenberg_2004}
\bibinfo{author}{\bibfnamefont{H.~S.} \bibnamefont{Eisenberg}},
  \bibinfo{author}{\bibfnamefont{G.}~\bibnamefont{Khoury}},
  \bibinfo{author}{\bibfnamefont{G.}~ \bibnamefont{Durkin}},
  \bibinfo{author}{\bibfnamefont{C.}~\bibnamefont{Simon}}, \bibnamefont{and}
  \bibinfo{author}{\bibfnamefont{D.}~\bibnamefont{Bouwmeester}},
  \bibinfo{journal}{Phys.\ Rev.\ Lett.} \textbf{\bibinfo{volume}{93}},
  \bibinfo{pages}{193901} (\bibinfo{year}{2004}).

\end{thebibliography}
\end{document}